# An Economic Perspective on Predictive Maintenance of Filtration Units

Tan Jing Yu, Denis
Nanyang Business School

Assoc Prof Law Wing-Keung, Adrian
School of Civil and Environmental Engineering

*Abstract* - This paper provides an economic perspective on the predictive maintenance of filtration units. The rise of predictive maintenance is possible due to the growing trend of industry 4.0 and the availability of inexpensive sensors. However, the adoption rate for predictive maintenance by companies remains low. The majority of companies are sticking to corrective and preventive maintenance. This is not due to a lack of information on the technical implementation of predictive maintenance, with an abundance of research papers on state-of-the-art machine learning algorithms that can be used effectively. The main issue is that most upper management has not yet been fully convinced of the idea of predictive maintenance. The economic value of the implementation has to be linked to the predictive maintenance program for better justification by the management.

In this study, three machine learning models were trained to demonstrate the economic value of predictive maintenance. Data was collected from a testbed located at the Singapore University of Technology and Design. The testbed closely resembles a real-world water treatment plant. A cost-benefit analysis coupled with Monte Carlo simulation was proposed. It provided a structured approach to document potential costs and savings by implementing a predictive maintenance program. The simulation incorporated real-world risk into a financial model. Financial figures were adapted from CITIC Envirotech Ltd, a leading membrane-based integrated environmental solutions provider. Two scenarios were used to elaborate on the economic values of predictive maintenance. Overall, this study seeks to bridge the gap between technical and business domains of predictive maintenance.

**Keywords** – industry 4.0, predictive maintenance, machine learning, filtration unit, business, proof of concept, cost-benefit analysis

## 1 INTRODUCTION

With the advancement of industry 4.0 and inexpensive sensors, predictive maintenance has become more than ever possible. Industry 4.0 seeks to fusion machines with data to improve overall performance and maintenance management [1]. This movement serves to enable real-time remote monitoring of physical assets, live tracking of products, and streamlining the entire production process [2]. Companies are increasingly aware of the benefits of predictive maintenance. Yet, the adoption rate remains low. According to Emory University Future of IIoT Research Study and Presenso, only 35% of industry players deploy advanced statistical modeling or machine learning for predictive analytics [3].

There exists an abundance of research papers purely focusing on the technical implementation of predictive maintenance [4, 5]. They provide detailed methods describing the use of statistical modeling, time series forecasting, or machine learning to predict faults. These are useful for companies that have already implemented predictive maintenance and are looking to advance further. However, as mentioned earlier, most companies have not even begun implementing predictive maintenance. Technical details are of lower importance to these companies. The more pressing issue at hand is to justify the economic value of a predictive maintenance program. Upgrading to a predictive maintenance program is a long-term plan and requires substantial costs. If the management does not buy into the idea due to lack of cost justifications, no amount of state-of-the-art predictive maintenance models will suffice.

The scope of this study encompasses the technical and business aspects of predictive maintenance implementation. The focus is on bridging the gap between the two aspects. The study assumes reference to the water treatment industry, specifically the ultra-filtration stage. The model development stage simply serves to illustrate the proof of concept and is not the main focus. There is no intention to promote any technical models here as there are plenty of such resources out there. A financial model is also provided to analyze the economic side of implementing a predictive maintenance program. A discussion will be made linking the technical and financial models.

The remaining of this paper is structured in the following manner. Section 2 discusses the literature review of common maintenance strategies and the prospects of predictive maintenance. Section 3 describes and explains the methods used in this study to bridge the gap between the technical and business aspects of predictive maintenance. Section 4 shows the results obtained. Section 5 discusses the implications of the results and the limitations of the study. Lastly, Section 6 concludes the study done and outlines possible future related work.

## 2 LITERATURE REVIEW

Maintenance of plant and equipment frequently occurs in the engineering industries. These maintenance operations usually take up around 15% to 70% of the companies' overall expenses [6]. In other words, the maintenance cost can be a significant part of the operation cost. In an increasingly competitive landscape, companies need to constantly find ways to cut costs or risk losing to their competitors. Currently, corrective maintenance and preventive maintenance are the dominant maintenance strategies adopted by most companies. However, these two techniques suffer from a lack of foresight [7]. When the equipment suffers from an unanticipated breakdown, the factory or plant has to halt the production process until the equipment is fixed. This negatively impacts the revenue of the company or even the risk of losing its client due to late deliveries.

Due to the weaknesses of corrective maintenance and preventive maintenance, they have given rise to predictive maintenance. Unlike the other two, predictive maintenance follows a proactive approach. Sensors are utilized to monitor the values of the physical assets and they are crucial to the implementation of predictive maintenance [8]. They detect and provide various information such as time, temperature, pressure, length, mass, angle, length. The values can be displayed locally at the sensors displays or sent to an external network for further analysis [9].

There are two well-known concepts regarding how sensors provide signals on faults [10]. The first concept involves looking at faults from a data-centric view. Essentially, the values from the sensors are used to provide signals on impending faults. There are 4 main faults. Outlier refers to an isolated data point based on other data points. Gradient [11] and distance from other readings [12] are popular measures to detect outliers. Spike refers to multiple data points forming an unexpected uptrend, usually within a short period. "Stuck-at" refers to data points with little to no variation for an extended period. This phenomenon can happen after a spike. High noise or variance refers to data points that vary wildly. The underlying trend of the data might not be discovered. This should be differentiated from the usual noise effect that is common to most sensor data. The second concept involves looking at faults from a system-centric view. Instead of focusing on the data, it examines the sensor and attempts to discover any physical malfunctions. The malfunctions are linked back to the data. Both the data-centric view and system-centric view are closely interlinked with each other. The 4 faults mentioned above can be detected through either concept or a combination of both.

Given that it is the era of big data and industry 4.0, the data-centric view has risen in popularity rapidly. Therefore, most predictive maintenance implementations are headed towards this path.

There are 4 distinct methods for fault detection which enable the data-centric approach. As discussed in [13], they are rule-based methods, estimation-based methods, time-series-based methods, and learning-based methods. Rule-based methods require extensive use of domain-specific knowledge to set constraints. For example, if the sensor reading rises above or below a specific threshold, a fault signal is produced. Estimation-based methods make use of multiple similar sensors that are correlated. As a result, any individual sensor measurements can be estimated using the other sensor measurements. Time-series-based methods assume the inherent non-randomness of data. It utilizes concepts such as stationarity and autocorrelation to model the sensor measurements. A comparison can be made between the actual values and the predicted values to detect any possible faults with the system. Learning-based methods use machine learning to predict continuous values or classify faults. The data is usually split into training data and validation data. With a sufficient amount of data to represent normal and abnormal behaviors, learning-based methods can provide promising results.

## 3 METHODS

This study is conducted using a proof of concept approach. There are 4 main steps involved in this study. The first step involves collecting data to facilitate the process of building a predictive maintenance model. The second step involves the machine learning model development process after data collection. The third step involves the creation of a financial model to perform financial analysis and simulations. The fourth step attempts to link the predictive maintenance model to the financial model. Details of each step are further outlined below.

### 3.1 EXPERIMENTS AND DATA COLLECTION

Experiments and collection of data were performed at the Secure Water Treatment (SWaT) testbed located at the Singapore University of Technology and Design (SUTD). The testbed provides a platform that simulates a real-world water treatment plant. It has a physical and cyber portion. In this study, the focus is on the physical portion. There are 6 stages in the physical process: raw water intake; chemical dosing pre-treatment; ultra-filtration; dechlorination using ultraviolet lamps; reverse osmosis; backwash process. The experiments mainly involved the ultra-filtration stage and simulations of sensor values.

In addition to the 6 stages of the physical water treatment plant, the experiment setup also involved the usage of Allen-Bradley Programmable Logic Controllers (PLCs), Human Machine Interfaces (HMIs), Supervisory Control and Data Acquisition (SCADA) workstation and a Historian. The SCADA system makes available the sensor data during the operation of the testbed. The Historian records the sensor data which can be retrieved by the researchers after the experiment.

An experiment started by switching on the testbed system. This initiated stage 1 of the water treatment process, i.e. the raw water intake stage. Under no interference from researchers, the entire system goes through normal operation. However, most researchers are not interested in normal operations. In this case, various attack scenarios can be performed using the HMI. Changes can also be made to the entire physical process by changing the PLC code. However, this is not done here. The focus of the experiments for this study was on the predictive maintenance scenarios through simulation of values. The system takes around 20 minutes to stabilize after switching on. The values should preferably be changed after the system stabilized to avoid excessive noise factor. For the present experiments, the value was changed for the differential pressure indicator transmitter (DPIT) sensor in the ultra-filtration stage. The alteration of the value was performed 5 times throughout the entire duration of the experiment. The sequence of value simulation is presented in Table 1.

Table 1 Sequence of value simulation

| Timestamp (d/m/yyyy h:mm) | Simulated value (Kilopascal, kPa) |
|---|---|
| 26/2/2020 15:44 | 35 |
| 26/2/2020 15:54 | 20 |
| 26/2/2020 16:05 | 35 |
| 26/2/2020 16:17 | 20 |
| 26/2/2020 16:34 | 40 |

The experiment lasted approximately 145 minutes. The final dataset was exported as a comma-separated values (CSV) file. The dataset consisted of 8701 rows and 82 columns excluding the headings.

## 3.2 PREDICTIVE MAINTENANCE MODELS DEVELOPMENT

With the comma-separated values file obtained, a few predictive maintenance models were developed thereafter. The key steps taken in the model development process are described here. Python (programming language) was used for the entire process. The initial step consisted of loading the CSV file as a DataFrame object for easier data manipulation. Data cleaning was then performed by checking and removing any null values. The CSV file contained some "Bad Input" values which were removed. The data types of columns were mostly changed to floating-point numbers except for the timestamp column. Exploratory data analysis followed thereafter. Time series plots were visualized for columns belonging to the ultra-filtration stage. The next step would be feature extraction. The differential pressure indicating transmitter (DPIT301) was chosen as the primary feature as it represented the key parameter changed during the experiments. Lag features would be extracted from this primary feature. Lag 5 to lag 30 were chosen (lag n meant an offset of n-seconds). Next, a time series split was performed to obtain multiple training and test sets. Figure 1 shows a visual representation. A total of 5 splits was used.

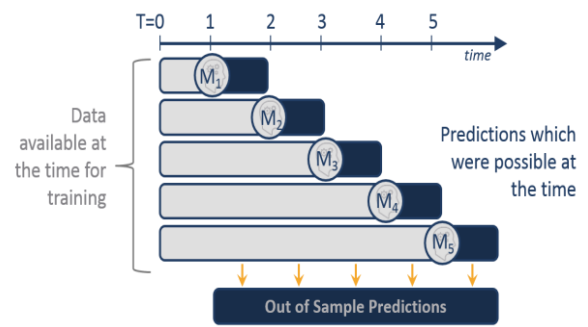

Figure 1 Time series split

Source: [14]

For the model building stage, 3 different types of models were used. They were linear regression, random forest regressor, and gradient boosting regressor. A model was trained once per time series split. This was repeated for all 3 regressors.

## 3.3 FINANCIAL MODELING

The financial modeling process seeks to create an abstract representation of an actual company's financial statement [15]. A cost-benefit analysis was chosen as the financial model to illustrate the proof of concept. The financial template used for the analysis mainly took reference from [16]. As mentioned earlier, the experiments were conducted in a testbed. Therefore, no financial figures could be obtained from the testbed itself. A possible solution was to come up with representative figures. However, it would be better to use figures from an actual company. This would help to achieve a more robust financial analysis and simulation. For this study, CITIC Envirotech Ltd was chosen as the company to be modeled after. It is a leading membrane-based integrated environmental solutions provider mainly engaged in environmental services. Most of the financial figures were adopted from its 2018 annual report [17]. Other figures were either obtained online or derived from existing figures in the annual report.

The cost-benefit analysis aims to serve as a non-biased assessment technique for the implementation of predictive maintenance programs. Ideally, the program manager can make decisions based on quantitative evidence backed by sound financial models. The template allows the manager to input any known potential costs and savings. Unquantifiable costs and savings should be clearly stated in the assumptions or notes. The core concept of the financial model is to present numbers accurately and transparently with reasonable assumptions.

Microsoft Excel was used to create the cost-benefit analysis template. The template utilized three spreadsheets. The first spreadsheet detailed the cost of implementing the predictive maintenance program. The second spreadsheet detailed the direct cost savings of the program. The third spreadsheet detailed the indirect cost savings of the program. All costs and savings are calculated based on an annual basis. The salient features of each spreadsheet are outlined below.

The first spreadsheet consists of 3 main categories: equipment, supplies/inventories, labor. The costs can be either variable only, fixed only, or both. The second spreadsheet details the reduction in costs attributed to operating activities and financing activities. A one-off gain due to disposal of equipment and supplies specific to preventive maintenance was also assumed. The third spreadsheet details the cost savings due to the avoidance of lost productivity and delay of maintenance cycles. Monte Carlo simulation was also utilized for values that are difficult to predict such as indirect cost savings and certain subcategories of direct cost savings. The simulation helped provide a probability distribution of outcomes and incorporated risk into the financial model. Refer to [18] for more details on the Monte Carlo simulation. The 3 spreadsheets are linked to model the net benefit or cost through the following equation:

Net benefit (cost) = Total direct cost savings + Total indirect cost savings − Total cost of implementation  (1)

It should be noted that the template is not meant to be a "once and for all" analysis. When hidden or latent costs surface throughout the predictive maintenance program, they should be reflected by updating the template. Companies can also expand upon this template to perform more complicated analyses such as return on investment or internal rate of return calculations.

## 3.4 LINKING PREDICTIVE MAINTENANCE MODEL TO FINANCIAL MODEL

Individually, the predictive maintenance model and the financial model are powerful tools already. However, this study is interested in bridging the gap between the technical and business aspects. To achieve this, two scenarios are designed to elaborate on the differences between the preventive maintenance and predictive maintenance.

The first scenario illustrates the avoidance of maintenance due to the predictive maintenance model signaling that no fault is likely to occur in the next X period. In contrast to the preventive maintenance cycle which initiates maintenance based on a fixed schedule, predictive maintenance can delay unnecessary maintenance which results in a sub-optimal operational time of the water treatment plant.

The second scenario illustrates the avoidance of unanticipated breakdown. The usual 30 minutes preventive maintenance cycle has the possibility of missing out on within-cycle failures. An unexpected event can cause the pressure of the filtration unit to spike up to an unacceptable range within a short period. If the predictive maintenance model can forecast this spike in pressure fast enough, the system can be halted swiftly. Thereafter, maintenance can be carried out. Catastrophic failure of the filtration unit is avoided as a result. By avoiding an extended period of downtime for the water treatment plant, significant loss of revenue can be prevented.

## 4 RESULTS

The trained machine learning models were evaluated using root mean square error (RMSE). RMSE is a standard statistical metric used to measure the standard deviation of the prediction errors. The formula for RMSE is defined as follow:

$$\text{RMSE} = \sqrt{\frac{1}{n} \sum_{i=1}^{n} (y_i - \hat{y})^2} \qquad (2)$$

where $n$ is the number of data points, $i$ is the index, $y_i$ is the actual value, and $\hat{y}$ is the predicted value. The evaluation results of the models on each test set are summarized in Table 2. The average RMSE of each model is also presented.

Table 2 Root mean square error results for all 3 models

| Split Number | Linear Regression RMSE (kPa) | Random Forest RMSE (kPa) | Gradient Boosting RMSE (kPa) |
|---|---|---|---|
| 1 | 3.46 | 5.07 | 5.06 |
| 2 | 1.41 | 2.85 | 2.68 |
| 3 | 2.61 | 3.67 | 3.38 |
| 4 | 2.45 | 3.38 | 18.29 |
| 5 | 0.74 | 2.19 | 0.91 |
| Average | 2.13 | 3.43 | 6.06 |

For the cost-benefit analysis, Monte Carlo simulation was evaluated to be suitable for 4 subcategories of both direct cost categories and indirect cost categories. The 4 subcategories are inspection cost savings, maintenance cost savings, avoidance of lost revenue, and materials cost savings. The results of the Monte Carlo simulation are presented in Table 3. The summary statistics shown include average, standard deviation, maximum, and minimum.

Table 3 Monte Carlo simulation for selected cost savings categories

|  | Average (S$'000,000) | SD (S$'000,000) | Max (S$'000,000) | Min (S$'000,000) |
|---|---|---|---|---|
| Inspection cost savings | 721 | 129 | 949 | 500 |
| Maintenance cost savings | 276 | 129 | 500 | 50 |
| Avoidance of lost revenue | 755 | 143 | 1000 | 500 |
| Materials cost savings | 6 | 3 | 10 | 1 |

## 5 DISCUSSION

From the results of the 3 machine learning models, linear regression performed the best based on the evaluation metric root mean square error. Random forest regressor was ranked second while the gradient boosting regressor ranked last. This was not unexpected as tree-based models are well-known for overfitting. Moreover, the data set was not that large with around 8700 rows. As this study focuses less on technical implementation, hyperparameter tuning was not performed. The comparison results might differ after hyperparameter tuning. A simple linear regression model has shown to be effective with the given data set. The advantages of these machine learning models can be linked back to the 2 scenarios mentioned in the methods section previously. With a reasonably low RMSE, the model can forecast the pressure values accurately. The manager can set a limit value for the sensor based on domain knowledge. For example, if the pressure value of the differential pressure indicating transmitter exceeds 40 kPa, maintenance actions have to be performed within the next 5 minutes. Otherwise, the water treatment plant can continue running. The delay of unnecessary maintenance would not only save costs but allow the plant to continue operating and generate additional revenue. The other scenario assumed the fact that equipment can fail rapidly due to unexpected events. The reactive nature of corrective maintenance and preventive maintenance would not be ideal here. Predictive maintenance would allow the timely detection of pressure value exceeding its limit and immediately halt the system to prevent further damage. Maintenance can be performed and the system continues operating thereafter. A significant loss of revenue can also be avoided. Generally, the model development process was satisfactory and served its purpose in illustrating the proof of concept.

The cost-benefit analysis financial template provided a structured way for managers to document all the potential costs and savings of the predictive maintenance program. Although it was specific to the water treatment industry, it can be easily generalized to any other similar industries. The template pushed for transparency. All assumptions or unknown factors could be documented down and made known to the management or potential investors. The nature of cost-benefit analysis was such that only relevant costs were included in the calculations. This meant sunk costs or any other unavoidable costs were excluded from the analysis. For example, if depreciation or research and development costs were included in the financial analysis, a profitable project might turn unprofitable. This would cause the predictive maintenance program to be rejected by the management without a doubt. Thus, cost-benefit analysis seeks to avoid such situations. It should be clear that performing this analysis does not guarantee that the predictive maintenance project would be justifiable based on cost savings. Assumptions are inherent to all financial models. The model is as good as the assumptions made. However, the structured approach allows managers to quickly decide on whether to accept or reject the project. Opportunity costs can be minimized as a result.

The Monte Carlo simulation was proposed as an added function to the cost-benefit analysis financial template. The probabilistic distribution generated by the simulation served various purposes. Firstly, it models the inherent risk that most or all real-world companies face. Few companies can judge from the onset whether a new project would be successful or not. The lack of certainty resulted in many prospective projects being rejected by upper management. With a structured financial template that outlines in detail the potential costs and savings, companies can gain better insights into the future profitability of a predictive maintenance program. Secondly, the simulation encapsulates the risk of machine learning models. No model is completely perfect and can predict all faults correctly. There are events where the filtration unit experiences a rapid rise in pressure but the model failed to detect the fault. As a result, the water treatment plant has to halt all current operations and fix the issue. The downtime experienced means a loss of revenue for the company. The expectation that any machine learning model can perform sub-optimally is incorporated into the cost-benefit analysis. Lastly, the presence of a probability distribution improves cost and savings transparency. A point estimate can be biased due to several reasons such as sampling bias or lack of detailed information. For this study, the lack of opportunity to obtain data from a real company resulted in greater uncertainty. Modeling a range of values would help avoid overly optimistic or pessimistic results.

There are a few limitations to this study. As it was conducted using a proof of concept approach, the findings are mostly theoretical. Therefore, reality could pan out very differently for some companies. Given that predictive maintenance programs are long-term in nature, a cost-benefit analysis might not be sufficient for large publicly listed companies. This is because these companies have high opportunity costs for funds. The implementation of a predictive maintenance program requires a huge outlay of funds in the first few years. These funds could have been used for other investment purposes. As such, a discounted cash flow (DCF) analysis would be more suitable for these companies. DCF analysis evaluates an investment today based on future cash flows. However, it is difficult for an outsider to obtain a company's required rate of return on investment. Complex financial analyses could not be made as a result. The data obtained from the testbed was not representative of an actual water treatment plant either. The physical process was extremely stable as it was not treating water for public or industrial use. The machine learning models could have presented overly optimistic results not easily achievable in an actual water treatment plant.

## 6 CONCLUSION AND FUTURE WORK

The proof of concept approach deployed in this study has shown the immense potential of predictive maintenance in the water treatment industry. The wider industry outlook for predictive maintenance is also very optimistic with the rise of industry 4.0 and the availability of inexpensive sensors. This study has successfully bridged the gap between the technical and business domains of predictive maintenance. A few machine learning models were deployed to illustrate the proof of concept. A cost-benefit analysis combined with Monte Carlo analysis has provided a structured approach to build a financial model that incorporates real-world risk. Possible future work includes collaborating with an actual water treatment company to obtain more realistic data and start a pilot program over there. A deep understanding of a business could lead to a more successful implementation of predictive maintenance. Overall, this study has provided an economic perspective on predictive maintenance of filtration units. The limits of predictive maintenance know no bounds and the concepts employed in this study can hopefully be extended to future studies as well.

## ACKNOWLEDGMENT

I would like to acknowledge the funding support from Nanyang Technological University – URECA Undergraduate Research Programme for this research project.

I would like to thank my supervisor, Assoc Prof Law Wing-Keung, Adrian for providing valuable guidance and framing the research direction throughout this project. I would also like to thank Liuyang, Yuying, and Giorgio for assisting in the data collection process and providing their insights on the project.